\documentclass[12pt,indent]{article}
\usepackage{a4,latexsym,amsmath,amssymb}
\usepackage[latin1]{inputenc}
\usepackage{float}
\usepackage{natbib}
\usepackage{graphics}
\usepackage[english]{babel}
\usepackage{tabularx}
\usepackage{lscape}
\usepackage{multirow}
\usepackage{rotating}
\usepackage{dsfont}
\usepackage{subfigure}
\usepackage{bbm}
\usepackage{wrapfig}
\usepackage{url}
\usepackage{booktabs}
\usepackage[table]{xcolor}
\usepackage{enumitem}
\usepackage{setspace}
\usepackage{calc}
\usepackage{ifthen}
\usepackage{framed}

\newcommand{\EV}{\operatorname{E}}

\newcommand{\blanco}[1]{}

\def\d{\displaystyle}

\newmuskip\pFqmuskip
\newcommand*\pFq[6][8]{%
  \begingroup 
  \pFqmuskip=#1mu\relax
  \mathcode`\,=\string"8000
  \begingroup\lccode`\~=`\,
  \lowercase{\endgroup\let~}\pFqcomma
  {}_{#2}F_{#3}{\left[\genfrac..{0pt}{}{#4}{#5};#6\right]}%
  \endgroup
}
\newcommand{\pFqcomma}{\mskip\pFqmuskip}

\usepackage{natbib}
\begin{document}
\bibliographystyle{chicago}
\sloppy

\makeatletter
\renewcommand{\section}{\@startsection{section}{1}{\z@}%
        {-3.5ex \@plus -1ex \@minus -.2ex}%
        {1.5ex \@plus.2ex}%
        {\reset@font\Large\sffamily}}
\renewcommand{\subsection}{\@startsection{subsection}{1}{\z@}%
        {-3.25ex \@plus -1ex \@minus -.2ex}%
        {1.1ex \@plus.2ex}%
        {\reset@font\large\sffamily\flushleft}}
\renewcommand{\subsubsection}{\@startsection{subsubsection}{1}{\z@}%
        {-3.25ex \@plus -1ex \@minus -.2ex}%
        {1.1ex \@plus.2ex}%
        {\reset@font\normalsize\sffamily\flushleft}}
\makeatother



\newsavebox{\tempbox}
\newlength{\linelength}
\setlength{\linelength}{\linewidth-10mm} \makeatletter
\renewcommand{\@makecaption}[2]
{
  \renewcommand{\baselinestretch}{1.1} \normalsize\small
  \vspace{5mm}
  \sbox{\tempbox}{#1: #2}
  \ifthenelse{\lengthtest{\wd\tempbox>\linelength}}
  {\noindent\hspace*{4mm}\parbox{\linewidth-10mm}{\sc#1: \sl#2\par}}
  {\begin{center}\sc#1: \sl#2\par\end{center}}
}


\def\R{\mathchoice{ \hbox{${\rm I}\!{\rm R}$} }
                   { \hbox{${\rm I}\!{\rm R}$} }
                   { \hbox{$ \scriptstyle  {\rm I}\!{\rm R}$} }
                   { \hbox{$ \scriptscriptstyle  {\rm I}\!{\rm R}$} }  }

\def\N{\mathchoice{ \hbox{${\rm I}\!{\rm N}$} }
                   { \hbox{${\rm I}\!{\rm N}$} }
                   { \hbox{$ \scriptstyle  {\rm I}\!{\rm N}$} }
                   { \hbox{$ \scriptscriptstyle  {\rm I}\!{\rm N}$} }  }

\def\d{\displaystyle}

\title{A Random Forest Approach for Modeling Bounded Outcomes}
\author{Leonie Weinhold, Matthias Schmid, \\Marvin N. Wright, Moritz Berger}

\maketitle

\begin{abstract}
\noindent Random forests have become an established tool for classification and regression, in particular in high-dimensional settings and in the presence of complex predictor-response relationships. For bounded outcome variables restricted to the unit interval, however, classical random forest approaches may severely suffer as they do not account for the heteroscedasticity in the data. A~random forest approach is proposed for relating beta distributed outcomes to explanatory variables. The approach explicitly makes use of the likelihood function of the beta distribution for the selection of splits during the tree-building procedure. In each iteration of the tree-building algorithm one chooses the combination of explanatory variable and splitting rule that maximizes the log-likelihood function of the beta distribution with the parameter estimates derived from the nodes of the currently built tree. Several simulation studies demonstrate the properties of the method and compare its performance to classical random forest approaches as well as to parametric regression models.   
\end{abstract}


\section{Introduction}\label{sec:intro}

In observational studies one frequently encounters bounded outcome variables. Important examples are (i) relative frequency measures restricted to the unit interval (0,1), like the household coverage rate \citep{fang2013}, the percentage of body fat \citep{fang2018} or DNA methylation levels \citep{weinhold2016}, and (ii) continuous response scales bounded between 0 and 100, like visual analogue scales \citep{bilcke2017} or health-related quality of life (HRQoL) scales~\citep{hunger2012}. 

The objective of statistical analyses typically is to build a regression model that relates the bounded outcome to a set of explanatory variables $X$~=~$(X_1,\hdots,X_p)^\top$. When modeling bounded outcomes a problem arises when the variance is not constant,
which implies heteroscedasticity and violates the assumption of Gaussian regression.
Therefore, a popular approach is to apply a transformation, e.g. by using the logistic function, that maps the unit interval (0,1) to $(-\infty,\,\infty)$ and to fit a Gaussian regression model to the transformed outcome. A remaining limitation of an analysis based on transformations is that inference is not possible on the original scale but only on the (logit-)transformed scale. This affects the interpretability of the estimated regression coefficients and the conclusions drawn from the results of associated hypothesis tests.

A more flexible method that does not require variable transformation is \textit{beta~regression}, see \citet{Ferrari.2004}, \citet{CribariNeto.2010} and the extensions considered by \citet{smithson2006}, \citet{simas2010} and \citet{grun2012}. One of the main benefits that makes the beta regression a popular tool for modeling bounded outcome variables is the flexibility of the shape of the probability density function (p.d.f.) of the beta distribution, including symmetrical shape and left or right skewness. Furthermore, beta regression allows for a simple interpretation of the predictor-response relationships in terms of multiplicative increase/decrease of the expected outcome, which is analogous to logistic regression for binary outcome. 

A large part of the methodology on beta regression refers to parametric regression models using simple \textit{linear} combinations of the explanatory variables, that is, one assumes that the effects of the explanatory variables on the outcome are linear. In many applications, however, this assumption is too restrictive, for example, when higher-order interactions between the explanatory
variables are present. Also, the specification of a parametric model may results in a large number of parameters relative to the sample size, which may effect estimation accuracy. When the number of explanatory variables exceeds the number of observations, parameter estimation is even infeasible. 

These issues can be addressed by the use of \textit{recursive partitioning techniques} or \textit{tree-based modeling}, which has its root in automatic interaction detection. The most popular version is due to \citet{Breiman1984} and is known by the name \textit{classification and regression trees} (CART). An easily accessible introduction into the basic concepts is found in \citet{Hastie.2009}. A disadvantage of tree-based methods, however, is that the resulting trees are often affected by a large variance, i.e.\,that minor changes in the data may result in
completely different trees. Therefore, in particular when the focus is on prediction accuracy, it is often worth stabilizing the results obtained from single trees by applying ensemble
methods, like \textit{random forests} \citep{breiman2001, ishwaran2007}. In general, the idea of random forests is to reduce the variance of the predictions while retaining low bias by averaging over many noisy trees. 

In this article, we propose a random forest approach tailored to the modeling of bounded outcome variables. In contrast to classical approaches, which use impurity measures (for classification) or the mean squared error (for regression), we propose to use the likelihood of the beta distribution as splitting criterion for tree building. In each iteration of the tree-building algorithm one chooses the combination of explanatory variable and split point that maximizes the log-likelihood function of the beta distribution, with the parameter estimates directly derived from the nodes of the currently built tree. 
Similar strategies have been proposed by \citet{su2004} for continuous outcomes and \citet{schlosser2018} in the context of distributional regression. 
The proposed method has been implemented in an extended version of the R add-on package \textbf{ranger}~\citep{ranger2017}.

The remainder of the article is organized as follows: in Section \ref{sec:methods} we present the notation and methodology of the proposed random forest approach for modeling bounded outcomes, along with a description of classical parametric beta regression models. Section \ref{sec:results} compares the properties and the performance of the proposed method to classical beta regression models and alternative random forest approaches considering several simulation studies. The main findings of the article are summarized in Section \ref{sec:conclusion}.

\section{Methods}\label{sec:methods}
\subsection{Notation and Methodology}\label{subsec:notation}
\begin{figure}[!t]
\centering
    \subfigure{\includegraphics[width=0.48\textwidth]{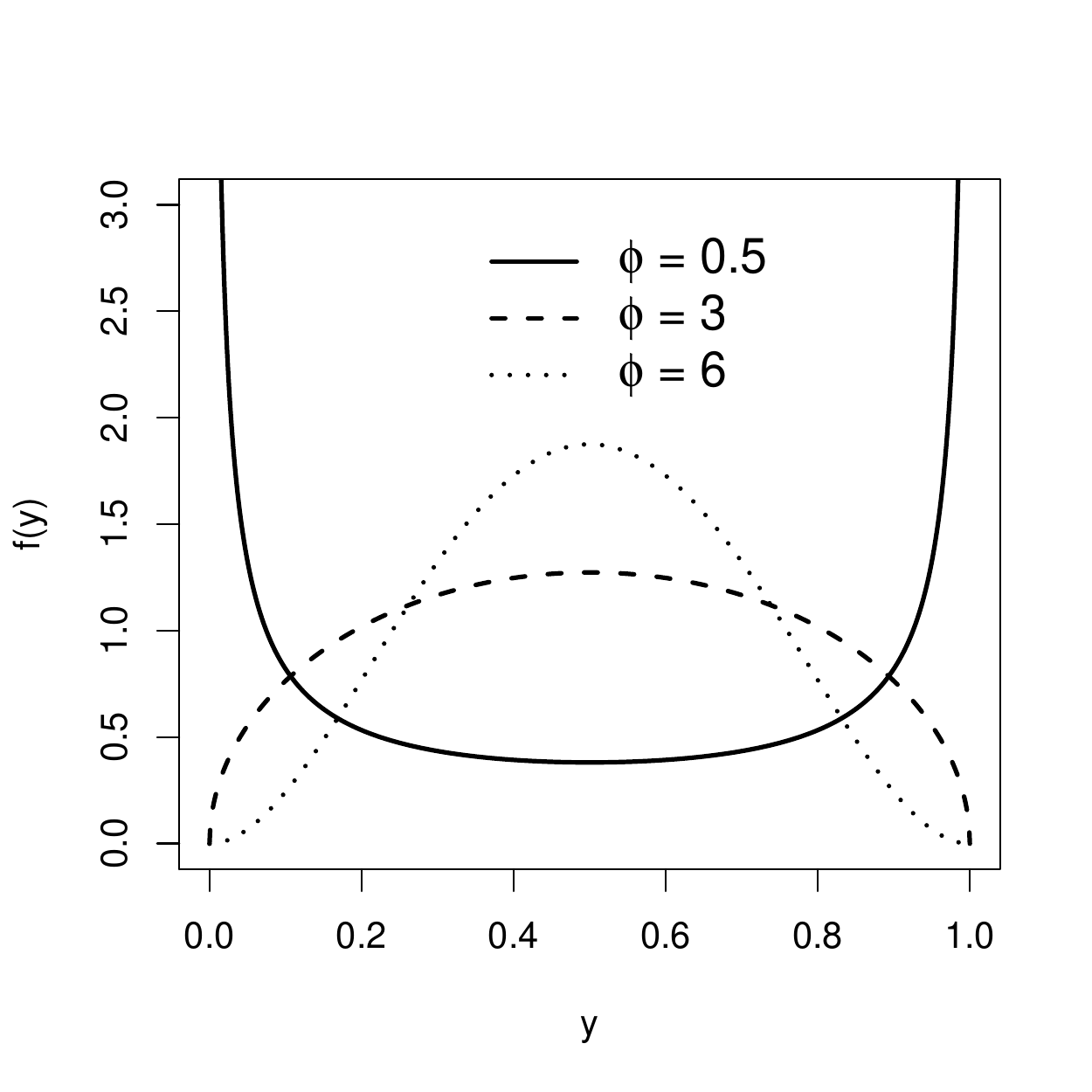}}
    \subfigure{\includegraphics[width=0.48\textwidth]{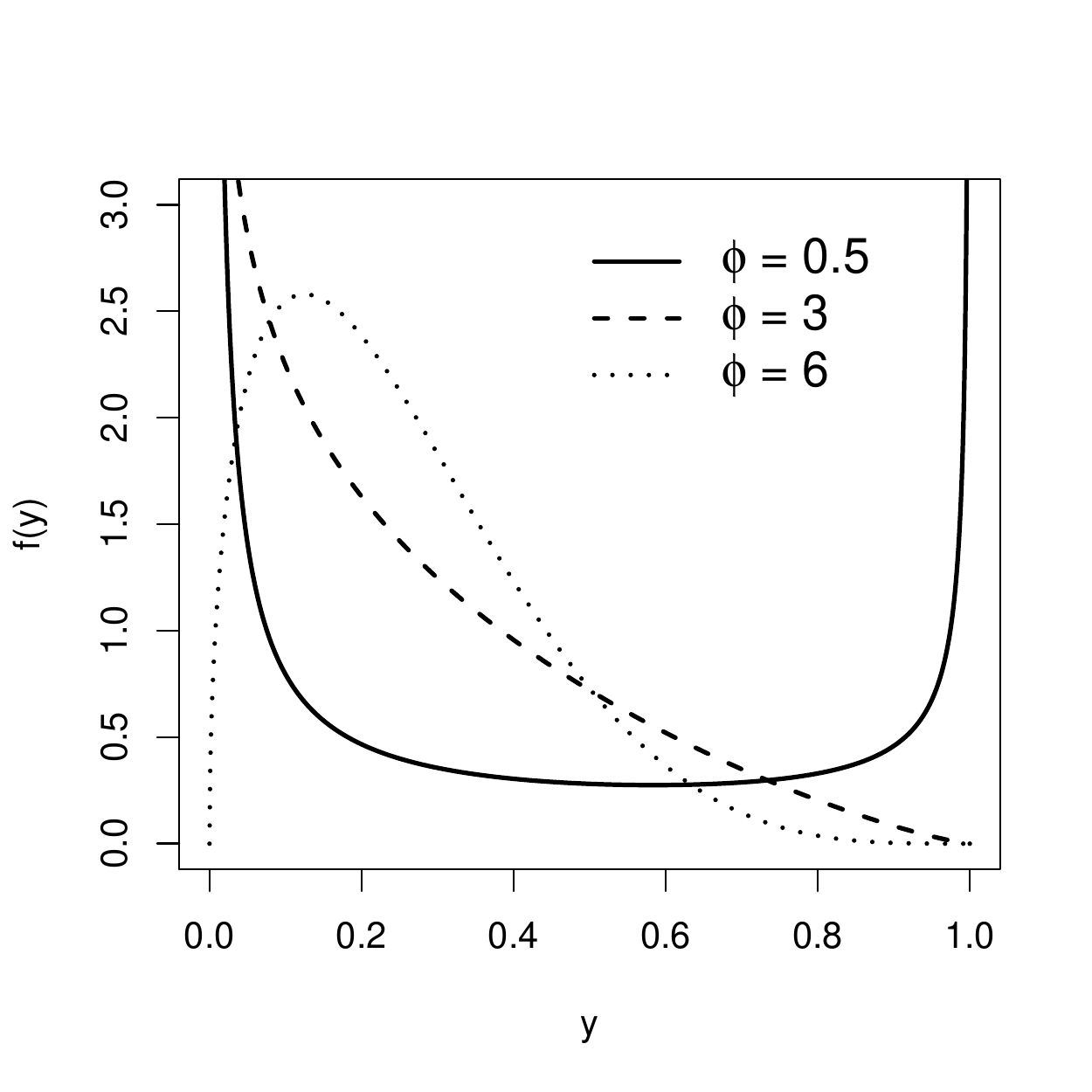}}
\caption{Examples of the probability density function of the beta distributed outcome $Y$ for parameter values $\phi \in \{0.5,3,6\}$ and fixed $\mu=0.5$ (left panel) and $\mu=0.25$ (right panel).}
	\label{fig:BetaDistributions}
\end{figure}

Let $Y \in (0,1)$ be the bounded outcome variable of interest. In this article we assume that $Y$ follows a beta distribution with the following probability density function 
	\begin{equation}
	\label{eq:betadensity}
	f_Y(y; \mu, \phi) =\, \frac{\Gamma (\phi) }{\Gamma (\mu \phi) \Gamma ((1-\mu)\phi )}\,
	y^{\mu\phi - 1}\, (1-y )^{(1-\mu )\phi - 1} \, ,
	\end{equation}
where $\Gamma(\cdot)$ is the gamma function. The p.d.f.\,in \eqref{eq:betadensity} is defined by two parameters, namely the location parameter $\mu \in (0,1)$ and the precision parameter $\, \phi>0$. With this parametrization, the mean and variance of $Y$ are, respectively, 
\begin{align}
\label{eq:MeanVarBetaDist}
	\EV(Y)= \mu \qquad \text{ and } 	\qquad \text{var}(Y) = \frac{\mu(1-\mu)}{\phi+1} \,.
\end{align}
Depending on the values of the parameters $\mu$ and $\phi$, the p.d.f.\,of the beta distribution can take on a number of different shapes, including symmetrical shape and left or right skewness, or even uniform shape \citep{CribariNeto.2010}. The flexibility is illustrated in Figure \ref{fig:BetaDistributions}, which shows the p.d.f.\,of $Y$ for several different parameter combinations.
Of note, the beta distribution can not only be used to characterize outcome variables bound to the unit interval $Y\in(0,1)$, but also for bounded variables of the type $\tilde{Y} \in (a,b)$ (with $a,b \in \mathbb{R}$, $a<b$) after scaling $\tilde{Y}$ to the standard unit interval by the transformation $(\tilde{Y}-a)/(b-a)$  \citep{CribariNeto.2010}. For further details on the beta distribution and alternative parameterizations, see \citet{Ferrari.2004}.

To relate $Y$ to the vector of explanatory variables $X=(X_1, \ldots, X_p)^T$, one usually considers the general class of parametric regression models and assumes that the relationship between the conditional mean $\mu|X:=\EV(Y|X)$ and $X$ is given by
\begin{align}
\label{eq:eta}
\mu|X = h(X^T\beta) \, = \, h(\eta(X)) \,,
\end{align}
where $h(\cdot)$ is a monotonic, twice differentiable response function, e.g. the inverse logit function, $\eta(\cdot)$ is the predictor function, and \mbox{$\beta=(\beta_1, \ldots, \beta_p)^T$} is a vector of unknown real-valued coefficients. When using the inverse logit function, the terms $\exp(\beta_1),\hdots,\exp(\beta_p)$ have a simple interpretation in terms of multiplicative increase or decrease of the conditional mean $\mu|X$. For example, if $\beta_k > 0, k \in \{1,\hdots,p\}$, increasing $X_k$ by one unit implies that the odds of the outcome \mbox{($\mu \,/\, 1-\mu$)} is increased by the factor $\exp(\beta_k)$. The model based on the inverse logit function will be used for comparison purposes in the simulation study in Section~\ref{sec:results}.

Estimates of the regression parameters $\beta$ are obtained by using classical maximum likelihood estimation. Let $y_i,\;i=1,\hdots,n,$ be a set of independent realizations of $Y$, with mean \mbox{$\mu_i$} and precision $\phi_i$ and let \mbox{$x_i= (x_{i1}, \ldots, x_{ip})^T,\;i=1,\hdots,n,$} be real-valued observations on $p$ explanatory variables.
Then, the log-likelihood function of model~\eqref{eq:eta} is defined as 
\begin{align}
\begin{aligned}
l(y_1, \ldots, y_n,&\,\mu_1, \ldots, \mu_n,\,\phi_1, \ldots, \phi_n) \\
=&\sum_{i=1}^n\bigg\{\log(\Gamma(\phi_i))  -\log (\Gamma(\mu_i\phi_i)) - \\
&\log(\Gamma((1-\mu_i)\phi_i))  + (\mu_i\phi_i-1)\log(y_i) + \\
&((1-\mu_i)\phi_i-1)\log(1-y_i)\bigg\} \,. 
\end{aligned} \label{eq:betaLL1}
\end{align}
According to Equation \eqref{eq:eta}, the log-likelihood function in \eqref{eq:betaLL1} can alternatively be written as a function of $x_i$ and $\beta$ (instead of $\mu_i$), given by 
\begin{align}
\begin{aligned}
l(y_1, \ldots, y_n,&\,x_1, \ldots, x_n, \beta,\,\phi_1, \ldots, \phi_n) \\
=&\sum_{i=1}^n \bigg\{\log(\Gamma(\phi_i))  -\log (\Gamma(h(x_i^T\beta)\phi_i)) - \\
&\log(\Gamma((1-h(x_i^T\beta))\phi_i)) + (h(x_i^T\beta)\phi_i-1)\log(y_i) + \\
&((1-h(x_i^T\beta))\phi_i-1)\log(1-y_i) \bigg\} \, .
\end{aligned} \label{eq:betaLL2}
\end{align}
Typically, in beta regression models, the precision parameter is assumed to be constant for all observations ($\phi_i \equiv \phi \;\text{for}\;i=1,\hdots,n$). This is the same as the assumption for the variance $\sigma^2$ in a Gaussian regression model. The more general form of the log-likelihood function specified in Equations \eqref{eq:betaLL1} and \eqref{eq:betaLL2} is exploit by the proposed approach (see Section \ref{subsec:regTree}).

The parametric model \eqref{eq:eta} is linear in $\beta$, implying that the explanatory variables have a linear effect on the transformed outcome. In practice, this assumption may be too restrictive, because explanatory variables may show complex interaction patterns. In principle the linear relation in \eqref{eq:eta} can be extended by incorporating interaction terms. However, for modeling interactions in classical regression approaches, the specification of the corresponding terms in the predictor function $\eta(X)$ of the model is required, i.e. they need to be known beforehand. This is a major challenge in real applications, due to the mostly unknown nature of these terms. 
Yet another problem when applying classical regression approaches arises when the number of parameters exceeds the number of observations (which is a likely scenario when many interaction terms are contained), i.e. when $p>n$. This leads to an overdetermined system for which maximum likelihood estimation is bound to fail. 

\subsection{Tree-Based Beta Regression}\label{subsec:regTree}

To address the aforementioned problems, several statistical methods were developed, including the popular Classification and Regression Trees (CART; \citealp{Breiman1984}). 
CART are built by recursively dividing the predictor space (defined by the explanatory variables X) into disjoint regions (i.e. subgroups of the observations) applying binary splits. 
Starting from the root of the tree (called \textit{top~node}), which contains the whole predictor space, in each splitting step, a single explanatory variable and a corresponding split point is selected, along which the node, denoted by $M$, is divided into two subsets $M_1$ and $M_2$ (called \textit{child nodes}). 
The scale of the variable $X_k,\,k=1,\hdots,p\,,$ defines the form of the binary splitting rule. For a metrically scaled or ordinal variable $X_k$, the partition into two subsets has the form $M_1 = M \cap \{X_k \leq c\}$ and  $M_2 = M \cap \{X_k > c\}$, with regard to split point~$c$. For multi-categorical variables without ordering $X_k \in \{1, \ldots,r\}$, the partition has the form  $M_1 = M \cap C_1$ and $M_2 = M \cap C_2$, where $C_1 \subset \{1,\ldots,r\}$ and $C_2 \subset \{1,\ldots,r\}$\textbackslash $C_1$ are disjoint, non-empty subsets. 
Splitting is repeated in each newly created child node until a specified stopping criterion based on predefined tuning parameters is met (see Section \ref{subsec:betaForests}). In each resulting terminal node, the conditional mean $\mu|X$ of all observations belonging to this node is fitted by a constant, e.g.\,the mean of the outcome values in regression trees and the most frequent class in classification trees. 

For classification trees, one of the most common splitting criteria (to choose among the explanatory variables and splitting rules in each step), is the \textit{Gini impurity}, which is minimal if all the observations are correctly classified. For regression trees, 
the usual splitting criterion is the \textit{mean squared error} (MSE). 
In the presence of a beta distributed outcome variable, one inherently assumes heteroscedasticity, because the variance depends on the mean parameter, see Equation~\eqref{eq:MeanVarBetaDist}. 
Hence, the use of the mean squared error as splitting criterion may lead to biased predictions, since splits are stronger affected by data with high variability. More specifically, because regression trees seek to minimize the within-node variance, there will be a tendency to split nodes with high variance, which may result in poor predictions in low-variance nodes \citep{moisen2008}. 

Therefore, we introduce an alternative splitting criterion based on the log-likelihood of the beta distribution, which forms the building block of the proposed random forest approach (see Section \ref{subsec:betaForests}). Given the data $(y_i,x_i),\,i=1,\hdots,n$, the log-likelihood \eqref{eq:betaLL1} can be obtained by estimating the distribution parameters $\mu_i$ and $\phi_i$ for each observation $i$. During the tree-building procedure, the estimation of the two parameters is conducted node-wise: in each node $M$ the location parameter $\mu_M$ and the precision parameter $\phi_M$ of the beta distribution are estimated by 
\begin{align}
\label{formula:betatreeMU}
\mu_{M} = \frac{1}{|M|}\,\sum_{i \in A_M}y_i
\qquad \text{and} \qquad 
\phi_{M} = \frac{\mu_M(1-\mu_{M})}{\frac{1}{|M|-1}\sum_{i \in A_M}(y_i-\mu_M)^2} -1 \, ,
\end{align}
where $A_M=\{i \in \{1, \ldots, n\}| x_{i} \in M\}$ is the set of observations and \mbox{$|M|=\sum_{i=1}^n\mathds{1}_{x_{i} \in M}$}  is the number of observations falling into node $M$, respectively. Given $Q$ nodes of a currently built tree, the individual distribution parameters for each observation $i$ are then obtained by  
\begin{align}
\label{formula:betatreeMU_obs}
&\mu_i=\mu_M, \,  \forall i \in A_M, \quad M=1,\hdots,Q, \, i=1,\ldots,n\,, \\ 
&\phi_i=\phi_M, \,  \forall i \in A_M , \quad M=1,\hdots,Q,\, i=1,\ldots,n\,. 
\end{align}
In each iteration of the proposed tree-building algorithm, one chooses the combination of explanatory variable $X_k$ and splitting rule, that maximizes the log-likelihood function \eqref{eq:betaLL1}, when node $M$ is split into the child nodes $M_1$ and $M_2$.
As in the classical tree approach, after termination of the algorithm the fitted conditional mean $\mu|X$ for each observations is obtained by Equation \eqref{formula:betatreeMU_obs}. 

\subsection{Random Forests for Beta Regression}\label{subsec:betaForests}

A great advantage of tree-based methods is the simple and intuitive interpretability of the model. This is particularly important when the aim is to build an easy-to-interpret prediction formula and to quantify the effect of a specific explanatory variable on the outcome. A drawback, however, is that the resulting tree estimators are often very unstable. This means that a small variation in the data can result in very different splits, which particularly affects prediction accuracy. 
To overcome this problem, we propose a random forest algorithm, an ensemble method originally developed by \citet{breiman2001}. The proposed approach is based on the splitting criterion introduced in Section \ref{subsec:regTree} and is referred to as \textit{beta forest} in the following.  

The main principle, of the beta forest is to generate a fixed number of samples from the original data (denoted by \textit{ntree}) using bootstrap procedures, i.e.\,by drawing with replacement, and to apply the tree-building algorithm to each of the samples. To mitigate the similarity of the resulting trees, the number of explanatory variables that are available for splitting in each node (denoted by \textit{mtry}) is reduced. During tree building, each node is split until the number of observations falls below a (predefined) minimal node size (denoted by \textit{min.node.size}). The algorithm finally  terminates when all current nodes are flagged as terminal nodes. As in the classical random forest approach, the fitted conditional mean~$\mu|X$ for an observation $i$ is obtained by averaging the predicted values over all trees where the observation was part of the \textit{out-of-bag} sample, i.e.\,was not used for tree building. A schematic overview of the beta forest algorithm is provided in Table~\ref{tab:betaforest}. 

\begin{table}[!t]
\caption{Schematic overview of the proposed beta forest. During tree building, the algorithm applies the splitting criterion introduced in Section \ref{subsec:regTree}.}
\begin{framed}
\begin{description}[leftmargin=3.4cm, style=multiline]
\item[\textbf{Initialization:}] Fix \textit{ntree}, \textit{mtry} and \textit{min.node.size}.
\item[\textbf{Bootstrapping:}] Draw \textit{ntree} bootstrap samples from the original data.
\item[\textbf{Tree Building:}] For each of the \textit{ntree} bootstrap samples fit the tree-based beta regression model (as described in \mbox{section \ref{subsec:regTree}}). More specifically, in each node of the trees, 
\begin{itemize}
\item[\bf{--}] draw \textit{mtry} candidate variables out of $p$ variables,
\item[\bf{--}] select the candidate variable and the splitting rule that maximize the log-likelihood of the beta distribution and split the data into two child nodes,
\item[\bf{--}] continue tree growing as long as the number of observations in each node is larger than \textit{min.node.size}. 
\end{itemize}
\item[\textbf{Prediction:}] For a new observation, drop it down to the final nodes of the \textit{ntree} trees built in step 'Tree Building'. Compute the ensemble estimate of the conditional mean by averaging the \textit{ntree} estimates of $\mu|X$. 
\end{description}
\end{framed}
\label{tab:betaforest}
\end{table} 

In R, the proposed algorithm has been implemented in an extended version of the add-on package \textbf{ranger}~\citep{ranger2017}. The default values of the tuning parameters of the beta forest were set to $ntree=500$, $mtry=\sqrt{p}$ and $min.node.size=5$, as for all the previously implemented regression methods.

\section{Results}\label{sec:results}

In this section we present the results of several simulations to evaluate the performance of the beta forest. The aims of the study were: (i) to compare the beta forest to classical random forest approaches in the presence of a bounded outcome variable, and (ii) to compare the beta forest to parametric beta regression models in higher dimensional settings with a large number of non-influential variables, and in the presence of interactions between the explanatory variables.

\subsection{Simulation Design}\label{subsec:SimulationData}

We considered several simulation scenarios, where each simulated data set consisted of $n_{train}=500$ observations. The outcome variable of each simulated data set was randomly drawn from a beta distribution with a constant precision parameter $\phi$ and the conditional mean $\mu|X = h(\eta(X_1, \ldots, X_4))$, with $h(\cdot)$ defined as the inverse logit function and the predictor function  
\begin{align}
\begin{aligned}\label{eq:predsim}
\eta(X_1, \ldots, X_4) = &\beta_0 +\beta_1\,X_1 + \beta_2\,X_2+\beta_3\,X_3+\beta_4\,X_4+ \\
&\beta_5\,X_1X_2 +\beta_6\,X_2X_3 + \beta_7\,X_3X_4 +\beta_8\,X_1X_4\,,
\end{aligned}
\end{align}
composed of four independent binary explanatory variables $X_k \in \{1,2\},$ with $P(X_k=1)=P(X_k=2)=0.5,\, k=1,\hdots,4$. Thus, the true predictor function~\eqref{eq:predsim} contained four linear main effects and four two-factor interactions. 

We simulated symmetrically distributed data with average location parameter value $\mu|X=0.5$ by setting $\beta=(0.2, 0.3,0.4,-0.1,-0.3, -0.3,-0.4,0.1, 0.3)^\top$, and left-skewed data with average location parameter value $\mu|X=0.8$ by setting \mbox{$\beta=(-0.2, -0.3,-0.4,0.1,0.3, 0.3, 0.4, 0.1, 0.3)^\top$}. For the precision parameter we used the values $\phi=2$ (\textit{low}), $\phi=4$ (\textit{moderate}) and $\phi=8$ (\textit{high}), yielding average variance parameter values of 0.08, 0.05 and 0.03 for $\mu|X=0.5$, and 0.05, 0.03 and 0.02 for $\mu|X=0.8$. Furthermore,  to assess the robustness of the methods against the degree of noise in the data we added non-informative explanatory variables $X_k \in \{1,2\}$ to the predictor space. Including the four informative variables, we considered scenarios with $p=4$ (\textit{informative}), $p=10$ (\textit{low-dimensional}), $p=100$ (\textit{moderate-dimensional}) and $p=200$ (\textit{high-dimensional}) explanatory variables. In total this resulted in \mbox{2 x 3 x 4 = 24} different scenarios. Each simulation scenario was replicated 100 times. 

\subsection{Competing Methods}

In all simulation scenarios we compared the beta forest (i) to alternative random forest approaches differing with regard to the transformation applied to the outcome variable $Y$, and (ii) to parametric beta regression models differing with regard to the pre-specified predictor function. Specifically, we examined:    

\begin{enumerate}[noitemsep]
  \item[(a)] the proposed beta forest (\textit{beta-rF}),
	\item[(b)]  the classical random forest approach with splitting criterion MSE on the original, untransformed data (\textit{rF}),
	\item[(c)]  the classical random forest approach with splitting criterion MSE on arcsine square root transformed data (\textit{asin-rF}),
	\item[(d)]  the classical random forest approach with splitting criterion MSE on logit-transformed data (\textit{logit-rF}),
	\item[(e)]  a fully specified parametric beta regression model with linear predictor~$\eta(X)=\beta_0+X^\top\beta$ (\textit{linear-bR}),
	\item[(f)] a parametric beta regression model without any explanatory variable, i.e.~predictor $\eta(X)=\beta_0$ (\textit{int-bR}) and 
	\item[(g)]  a parametric beta regression model with predictor function \eqref{eq:predsim} of the underlying data generating process  (\textit{true-bR}).
\end{enumerate}
All random forest approaches (a) -- (d) were computed with the R package \textbf{ranger} \citep{ranger2017} using the default values for \textit{ntree} and \textit{mtry} and $\textit{min.node.size}=10$. The parametric beta regression models (e) -- (g) were computed using the R package \textbf{betareg} \citep{CribariNeto.2010}. 

We evaluated the performance of the modeling approaches with respect to predicting outcomes of future observations. This was done by computing the predictive log-likelihood values (based on the beta, Gaussian, logit-normal or arcsine-normal distribution) on an independently drawn test data set of \mbox{$n_{test}=500$} observations in each replication.
To obtain the precision parameter $\phi$ of model (a), the log-likelihood function of the beta distribution was optimized on the sample of $n_{train}=500$ observations, after plugging in the conditional mean estimates of the random forest. The variance parameters $\sigma^2$ of the models (b) -- (d) were obtained accordingly.

\subsection{Comparison to Parametric Beta Regression}

\begin{figure}[!th]
	\centering
		\includegraphics[width=1\textwidth]{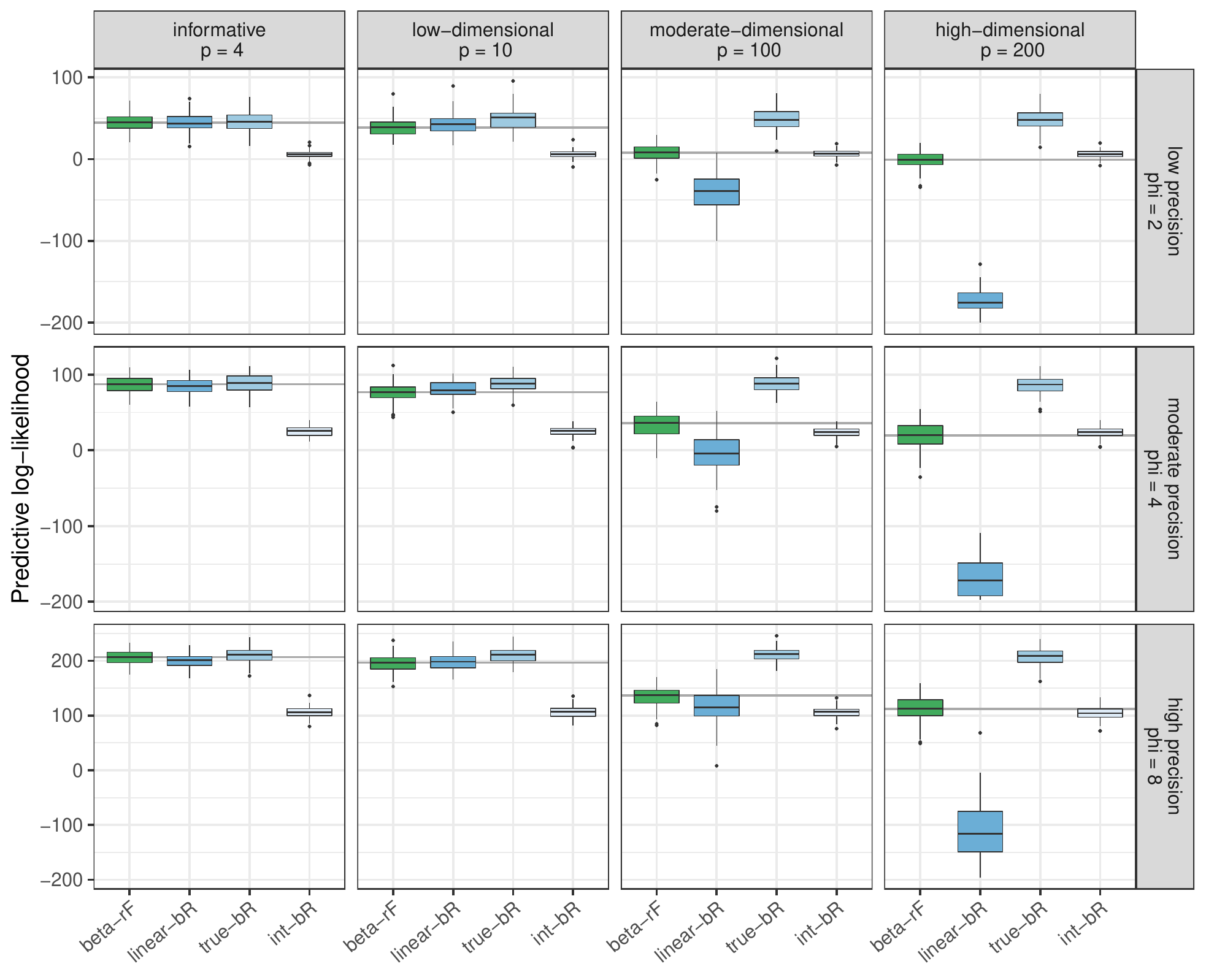}
	\caption{Predictive log-likelihood values of the parametric beta regression models and the beta forest for the 12 scenarios with symmetrically distributed outcome (average location parameter $\mu=0.5$). The precision parameter $\phi$ varies across the rows, the number of variables $p$ varies across the columns. The gray lines refer to the median values of the beta forest.}
	\label{fig:Results_symmetrisch_binaryLinearInt_beta}
\end{figure}

The prediction accuracy obtained from the beta forest and the parametric beta regression models (e) -- (g) for the 12 scenarios with symmetric outcome (average $\mu=0.5$) and varying $\phi$ (rows) and $p$ (columns) are shown in Figure \ref{fig:Results_symmetrisch_binaryLinearInt_beta}.

It is seen that the performance of the fully specified beta regression model~(linear-bR)  was comparable to the beta forest in the scenarios with only informative explanatory variables as well as in the scenarios with moderate-dimensional data (first and second column), with a slight advantage of the beta forest in the informative, high precision scenario ($p=4$, $\phi=8$). Further, the performance of the linear-bR substantially deteriorated with an increasing number of non-informative variables (third and fourth column of Figure \ref{fig:Results_symmetrisch_binaryLinearInt_beta}), strongly favoring the proposed beta forest. The difference becomes even more apparent for smaller values of the precision parameter $\phi$. For example in the moderate-dimensional, low precision scenario ($p=100$, $\phi=2$) the average difference in the predictive log-likelihood was 47.3. As was to be expected and throughout all settings, the parametric beta regression model with the predictor specified according to the true data-generating process~(true-bR) outperformed all other models. The relatively good performance of the simple int-bR model in the high-dimensional scenarios revealed that the explained variance in these data sets was rather small. For example, the average simulated $R^2$ was $0.23$ in the high-dimensional, moderate precision scenario ($p=200$, $\phi=4$). 

The results for the 12 scenarios with left-skewed data (average $\mu=0.8$) are shown in Figure \ref{fig:Results_linksschief_binaryLinearInt_beta} in Appendix \ref{sec:A1}. There are only minor differences to the previous results throughout all scenarios. 

The superiority of the beta forest in the moderate- and high-dimensional scenarios is mainly explained by the variable selection mechanism, which is enforced when fitting random forests. Moreover, the result in the informative, high precision scenario (where the performance is almost equal to the true-bR) stresses that the beta forest adequately accounted for the interaction terms contained in the data-generating model \eqref{eq:predsim}.

\subsection{Comparison to Random Forest Approaches}

\begin{figure}[!t]
	\centering
		\includegraphics[width=1.00\textwidth]{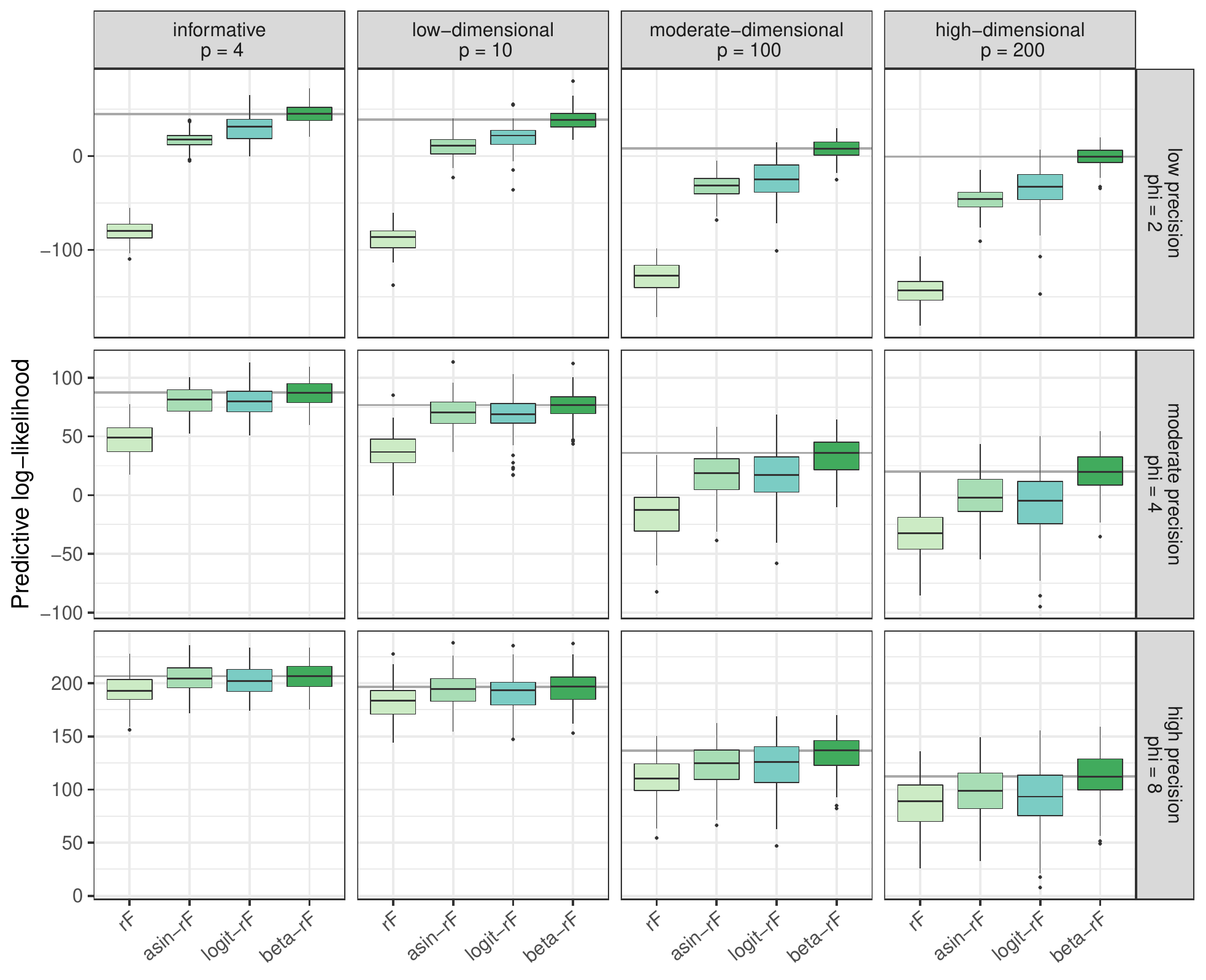}
	\caption{Predictive log-likelihood values of the random forest approaches for the 12 scenarios with symmetrically distributed outcome (average location parameter $\mu=0.5$). The precision parameter $\phi$ varies across the rows, the number of variables $p$ varies across the columns. The gray lines refer to the median values of the beta forest.}
	\label{fig:Results_symmetrisch_binaryLinearInt_rF}
\end{figure}

Figure \ref{fig:Results_symmetrisch_binaryLinearInt_rF} shows the prediction acuracy of the random forest approaches (a) -- (d) for the 12 scenarios with symmetric outcome (average $\mu=0.5$). Throughout all scenarios, the beta forest outperformed all competing random forest approaches.
Generally, all approaches gained in prediction accuracy with an increasing value of the precision parameter $\phi$ and a decreasing number of non-informative variables. The worst performance was obtained for the classical random forest on the untransformed data (rF), even though the difference vanished with increasing precision~$\phi$. The models based on arcsine square root transformed data (asin-rF) and logit transformed data (logit-rF) performed roughly equal in all scenarios. The largest differences to the beta forest were seen in the four low precision scenarios $\phi=2$ (first row of Figure \ref{fig:Results_symmetrisch_binaryLinearInt_rF}). For example, in the moderate-dimensional, low precision scenario, the average difference between logit-rF and beta-rF was~$33.1$. 

The results for the 12 scenarios with left skewed data (average $\mu=0.8$) are shown in Figure \ref{fig:Results_linksschief_binaryLinearInt_rF} in Appendix \ref{sec:A1}. One notable difference in these scenarios was the prediction performance of the logit-rF: in 
the low precision scenarios (first row of Figure \ref{fig:Results_linksschief_binaryLinearInt_rF}), the logit-rF yielded comparable results to the beta-rF and even outperformed the beta-rF for in the scenario with only informative explanatory variables and the low-dimensional data. Furthermore, the performance of the logit-rF strongly deviated from the asin-rF in these scenarios. 

The largely observed superiority of the beta forest over the competing random forest approach revealed that the beta forest was best able to account for the bounded structure of the outcome variable, which was simulated from a beta distribution.

\section{Concluding Remarks}\label{sec:conclusion}
We proposed a random forest approach for modeling bounded outcome variables. 
The beta forest is an alternative modeling strategy to parametric models if one finds to struggle with the specification of predictor-response relationships in higher dimensional settings. Furthermore, the random forest algorithm provides variable importance measures, which can be used to rank the explanatory variables in terms of their effect on the outcome variable.

The simulation study showed that (i) the beta forest yielded more accurate predictions of bounded outcomes than classical random forest approach based on the MSE as splitting criterion, and (ii) outperformed classical parametric beta regression models in the presence of high-dimensional data, as the method is capable to account for complex interaction patterns and carries out variable selection.  

The beta forest is implemented in the user-friendly R add-on package \textbf{ranger}~\citep{ranger2017} and can therefore easily applied by practitioners.

\bibliography{literatur}

\appendix 
\newpage
\section{Further Simulation Results} \label{sec:A1}

\begin{figure}[!h]
	\centering
		\includegraphics[width=1\textwidth]{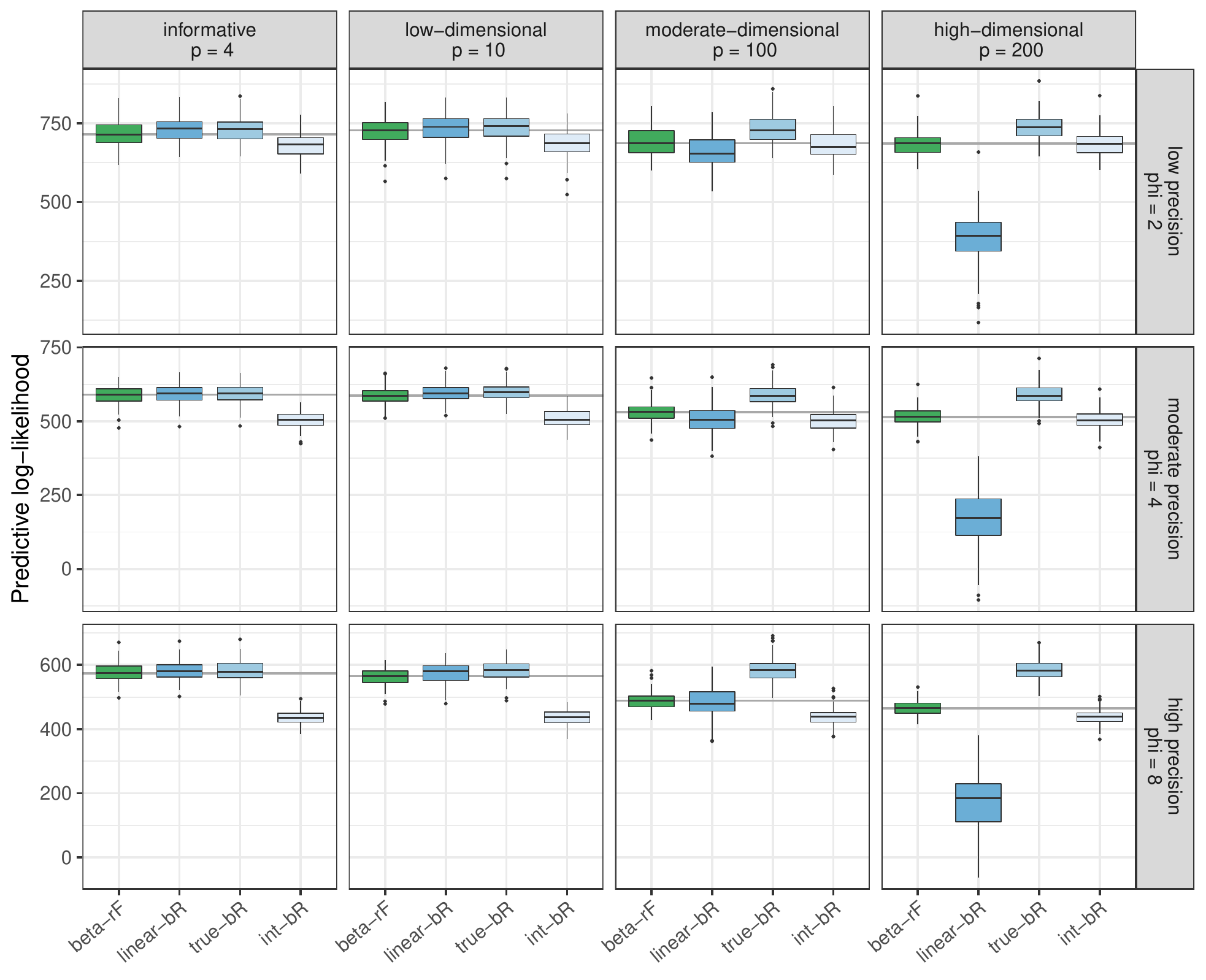}
	\caption{Predictive log-likelihood values of the parametric beta regression models and the beta forest for the 12 scenarios with left-skewed distributed outcome (average location parameter $\mu=0.8$). The precision parameter $\phi$ varies across the rows, the number of variables $p$ varies across the columns. The gray lines refer to the median values of the beta forest.}
	\label{fig:Results_linksschief_binaryLinearInt_beta}
\end{figure}

\begin{figure}[!h]
	\centering
		\includegraphics[width=1.00\textwidth]{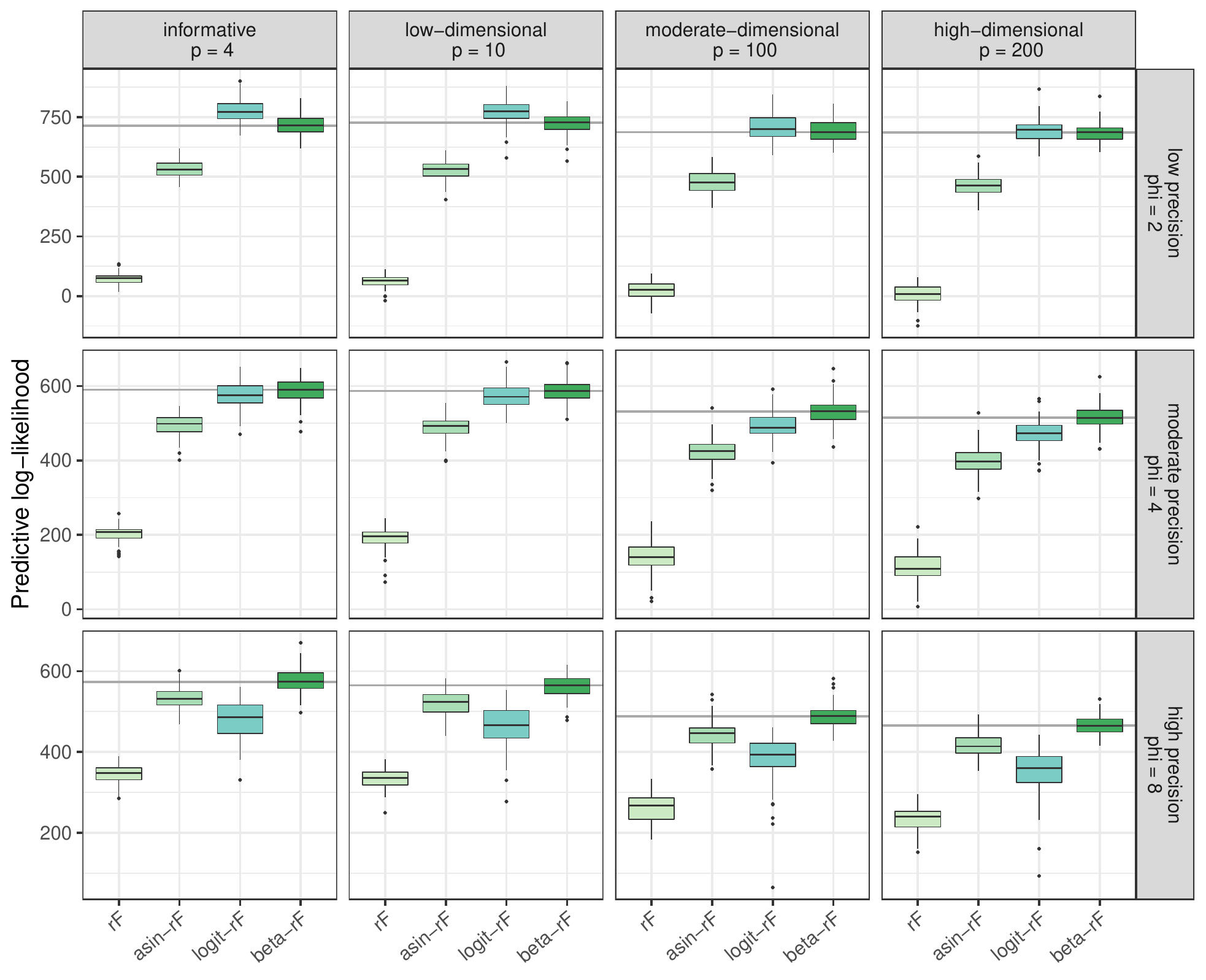}
	\caption{Predictive log-likelihood values of the random forest approaches for the 12 scenarios with left-skewed distributed outcome (average location parameter $\mu=0.8$). The precision parameter $\phi$ varies across the rows, the number of variables $p$ varies across the columns. The gray lines refer to the median values of the beta forest.}
	\label{fig:Results_linksschief_binaryLinearInt_rF}
\end{figure}

\end{document}